# Transmission lines emulating moving media

**New Journal of Physics**

The open access journal at the forefront of physics

Deutsche Physikalische Gesellschaft **DPG**  |  **IOP** Institute of Physics# Transmission lines emulating moving media

J Vehmas[1], S Hrabar[2] and S Tretyakov[1]

[1] Department of Radio Science and Engineering/SMARAD Center of Excellence, Aalto University, PO Box 13000, FI-00076 Aalto, Finland
[2] Faculty of Electrical Engineering and Computing, University of Zagreb, Unska 3, 10000 Zagreb, Croatia
E-mail: joni.vehmas@aalto.fiReceived 28 May 2014, revised 8 August 2014
Accepted for publication 27 August 2014
Published 30 September 2014

*New Journal of Physics* **16** (2014) 093065
doi:10.1088/1367-2630/16/9/093065## Abstract

In this paper, we show how the electromagnetic phenomena in moving magnetodielectric media can be emulated using artificial composite structures at rest. In particular, we introduce nonreciprocal periodically loaded transmission lines, which support waves obeying the same rules as plane electromagnetic waves in moving media. Because the actual physical structure is at rest, in these transmission lines there are no fundamental limitations on the velocity values, which may take values larger than the speed of light or even complex values (considering complex amplitudes in the time-harmonic regime). An example circuit of a unit cell of a 'moving' transmission line is presented and analyzed both numerically and experimentally. The special case of a composite right-/left-handed host line is also studied numerically. Besides the fundamental interest, the study is relevant for potential applications in realizing engineered materials for various transformations of electromagnetic fields.

Keywords: electromagnetics, moving media, metamaterial, transmission line, nonreciprocity## 1. Introduction

Electromagnetic fields in uniformly moving media can be studied, using the constitutive relations in two reference frames, moving with respect to each other. If in one reference frame

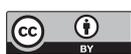 Content from this work may be used under the terms of the Creative Commons Attribution 3.0 licence.
Any further distribution of this work must maintain attribution to the author(s) and the title of the work, journal citation and DOI.*New Journal of Physics* **16** (2014) 093065
1367-2630/14/093065+20$33.00                                     © 2014 IOP Publishing Ltd and Deutsche Physikalische Gesellschaft

the medium appears to behave as an isotropic magneto-dielectric (permittivity $\epsilon'$, permeability $\mu'$, and square refractive index $n'^2 = c^2 \epsilon' \mu'$ with $c$ being the speed of light in vacuum), then in the reference frame which is moving with velocity $v$ along the axis $z$ ($\mathbf{v} = v\mathbf{z}_0$), the constitutive relations for the same medium take the form (e.g., [1, 2])

$$\mathbf{D}_t = \epsilon \mathbf{E}_t + \frac{V}{c} \mathbf{z}_0 \times \mathbf{H} \tag{1}$$

$$\mathbf{B}_t = \mu \mathbf{H}_t - \frac{V}{c} \mathbf{z}_0 \times \mathbf{E}, \tag{2}$$

where $\epsilon$ and $\mu$ are the effective permittivity and permeability of the medium, respectively, $V$ is a unitless velocity parameter, and $\mathbf{z}_0$ is the unit vector along $z$. Also, the index 't' denotes the field component transverse to the velocity direction $\mathbf{z}_0$: $\mathbf{E}_t = (\overline{\overline{I}} - \mathbf{z}_0 \mathbf{z}_0) \cdot \mathbf{E}$, where $\overline{\overline{I}}$ is the unit dyadic. The longitudinal field components are the same in both reference frames. The effective material parameters in moving media ($\epsilon$, $\mu$, and $V$) relate to the parameters at rest ($\epsilon'$ and $\mu'$) and the velocity of the moving frame or, alternatively, the velocity of the medium, depending on the viewpoint ($v$) by

$$\epsilon = \epsilon' \frac{1 - \frac{v^2}{c^2}}{1 - \frac{n'^2 v^2}{c^2}}, \qquad \mu = \mu' \frac{1 - \frac{v^2}{c^2}}{1 - \frac{n'^2 v^2}{c^2}}, \tag{3}$$

$$V = \frac{n'^2 - 1}{1 - \frac{n'^2 v^2}{c^2}} \frac{v}{c}. \tag{4}$$

These material relations have been used by many researchers to solve electromagnetic problems where more than two relatively moving media are involved. Obviously, in these relations the velocity value $v$ is always smaller than the speed of light $c$. It is important to point out that in the above analysis, the medium is assumed to be unbounded. If a finite slab of a moving medium is considered, the Doppler shift due to the movement of the interface between the two media must also be taken into account. The same is true for the case when a receiver and a transmitter are in relative motion.

On the other hand, considering (1) and (2), one notices that these are the constitutive relations of a bi-anisotropic medium [2, 3], which may potentially be realized as a composite material at rest. These material relations with anti-symmetric coupling dyadics correspond to a nonreciprocal class of bi-anisotropic media called, by analogy, *moving media* [3, 4], although the medium does not move, and the coupling is provided (due to its microstructure containing some nonreciprocal elements). It appears that materials obeying (1) and (2) have not been found among natural materials, but there have been conceptual suggestions on how such materials can be possibly realized using composites containing magnetized ferrite and metal inclusions [3, 5]. In the artificial-medium scenario, the values of the velocity parameter in the constitutive relations are not restricted to real values smaller than the speed of light, so this opens exciting opportunities to realize structures that emulate (fictitious) electromagnetic phenomena for media moving with superluminal or even complex-valued velocities.

Recently, the concept of artificial moving media was discussed in [6], where it was shown that an artificial moving medium transforms arbitrary electromagnetic fields, modulating them with an exponential function of the coordinate $z$ (the velocity direction). Real values of the 'effective' velocity correspond to an exponent with an imaginary argument, while the imaginary velocity corresponds to exponential decay or amplification of fields depending on the propagation direction. In [7], it was shown analytically that a thin layer (a single layer of hypothetical 'moving molecules') of an artificial moving medium can be tuned to work as an ideal isolator. Thus, realization of materials with the properties of moving media has not only fundamental theoretical interest but also would offer new application possibilities in, for example, absorbers and devices for manipulating the field distribution.

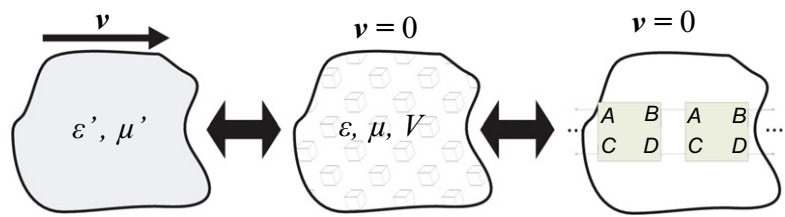

**Figure 1.** Connection between moving magnetodielectric medium (left), artificial moving medium (middle), and artificial TL moving medium (right).

Previously proposed structures having nonreciprocal bi-anisotropic responses have several limitations. Suggested realizations of an artificial moving medium based on embedding electrically small inclusions (consisting of a swastika-shaped metal element on top of a small magnetized ferrite sphere into a dielectric [3, 5]) suffer from parasitic chirality of the inclusions. Also, the suggested inclusions are resonant structures leading to a highly dispersive response. General nonreciprocal metamaterials have been considered in [8], where a nonreciprocal active metamaterial slab was realized. The slab was shown to be transparent to microwaves propagating in one direction and opaque in the opposite direction within a fairly narrow bandwidth. The bandwidth was limited in this case, again, due to the resonant nature of the inclusions comprising the slab. Furthermore, the proposed metamaterial slab was shown to exhibit a combination of moving and omega types of bi-anisotropic coupling instead of a pure moving medium response. Therefore, no practical realization of a moving molecule is known.

Here, we consider possibilities of realizing artificial moving media using periodically loaded transmission lines (TLs). First, we derive required conditions for moving media response in an infinite cascade of TL unit cells as well as the relations connecting the ABCD (transmission) parameters of a general periodic structure and the material parameters of moving media. Second, a circuit topology utilizing a gyrator is considered to fulfill the required conditions. Third, the introduced concept is validated experimentally. Finally, we study numerically the special case of the proposed circuit topology where the host line is a so called composite right-/left- handed TL. The concepts discussed here can, in principle, be applied to any frequency range though practical realizations at high frequencies may prove difficult. However, in this paper we limit the analysis to a low RF range, where experimental prototypes are easier to realize. The connection between a moving magnetodielectric medium, an artificial moving medium, and a suggested artificial TL moving medium is illustrated in figure 1. A similar study for another special case of bi-anisotropic media, the omega media, was conducted in [9].

## 2. Propagation constants and wave impedance of moving media

In view of emulating moving media by periodically loaded transmission lines, we will consider the constitutive equations (1)–(2) in frequency domain (the time-harmonic time dependence is in the form $\exp(i\omega t)$). The propagation constants $\beta_{\pm}$ for axially (along $z$) propagating plane waves in moving media can be easily derived, combining the Maxwell equations with the material parameters (e.g., [2, 5]) It is given by

$$\beta_{\pm} = \pm k_0(n \pm V), \tag{5}$$

where $k_0$ is the free-space wave number and $n = c\sqrt{\epsilon\mu}$ is the refractive index of the medium. The ± signs correspond to the opposite propagation directions. For moving media, depending on whether the incident wave propagates along the direction of the velocity vector ($+z_0$) or against it ($-z_0$), we get two different solutions for the propagation constant, as can be expected for a nonreciprocal medium. It is critically important that *both* eigenwaves decay or grow along the same fixed direction (positive or negative direction of the $z$ axis, depending on the sign of $V$; see [6]). The medium described by (1)–(2) is lossless if all the parameters $\epsilon$, $\mu$, and $V$ are real numbers (e.g., [3]). In this case, as it should be, the propagation constants are real numbers, and the eigenwaves do not decay. For a purely imaginary velocity $V$, the plane waves in the medium with the refractive index $n$ are modulated by the exponential function of a real argument. Polarization of the eigenwaves is linear, as in isotropic media. The wave impedance in such a medium is given simply by

$$\eta = \sqrt{\frac{\mu}{\epsilon}}, \tag{6}$$

where $\epsilon$ and $\mu$ are the permittivity and permeability of the medium, meaning that the wave impedance is independent of the velocity parameter $V$.

## 3. Required conditions for the TL unit cell

Now, we would like to equate the propagation constant and the characteristic impedance in the general cascade of periodically arranged and connected unit cells to the propagation constant and wave impedance in moving media, respectively. The unit cell of the cascade is characterized by its ABCD (transmission) parameters. These are defined according to [10]

$$\begin{bmatrix} V_1 \\ I_1 \end{bmatrix} = \begin{bmatrix} A & B \\ C & D \end{bmatrix} \begin{bmatrix} V_2 \\ I_2 \end{bmatrix}, \tag{7}$$

where $V_1$ and $I_1$ are the input voltage and current, respectively, while $V_2$ and $I_2$ are the corresponding quantities at the output. It is important to note that in this definition, it is assumed that the currents $I_1$ and $I_2$ are flowing in the same direction, from port 1 to port 2. The dispersion relation for an arbitrary periodic cascade of unit cells characterized by ABCD parameters can be easily derived using the definition of ABCD parameters (7) and the Floquet theorem [10] and has the form

$$\beta_\pm = -\frac{i}{d} \ln\left( \frac{A + D \pm \sqrt{(A+D)^2 - 4(AD - BC)}}{2} \right), \quad (8)$$

where ± signs, again, correspond to the opposite propagation directions and $d$ is the period of the cascade. The Bloch impedance can be considered as the characteristic impedance of periodically loaded transmission lines. It is defined as the ratio of the voltage and current at the terminals of the unit cell in an infinitely long cascade of such unit cells. It should be noted that the value of the Bloch impedance depends on how the terminal points are chosen and is, therefore, not unique for a given unit cell. The Bloch impedance for an arbitrary unit cell characterized by its ABCD parameters can be derived, again, using the definition of the ABCD parameters (7) and the Floquet theorem [10] and has, in the general case, the form

$$Z_B = \pm \frac{B}{A - e^{i\beta_\pm d}}. \quad (9)$$

Here, the currents are defined to point into the direction of wave propagation (instead of always in the same direction). This definition was chosen here, as it gives a single Bloch impedance with a real and positive value for a conventional unloaded TL. The impedance of moving media (6) is independent of the propagation direction, meaning that moving media are symmetric when it comes to impedance. This means that the Bloch impedances of the artificial unit cell for different propagation directions should also have the same value. By plugging (8) into (9), we can see that, in this case, we must have $A = D$. Taking this into account, (8) simplifies to

$$\beta_\pm = -\frac{i}{d} \ln(A \pm \sqrt{BC}). \quad (10)$$

By plugging (10) into (9), the Bloch impedance simplifies to

$$Z_B = \sqrt{\frac{B}{C}}, \quad (11)$$

which, as can be expected, is independent of the propagation direction.

By comparing the dispersion and wave impedance in moving media, (5) and (6), to the dispersion and Bloch impedance in a general chain characterized by ABCD parameters, (10) and (11), we can now solve the effective refractive index of the periodically loaded TL

$$n = -\frac{i}{2k_0 d} \ln\left( \frac{A + \sqrt{BC}}{A - \sqrt{BC}} \right). \quad (12)$$

Similarly, we can also solve the effective normalized velocity

$$V = -\frac{i}{2k_0 d} \ln\left( A^2 - BC \right). \quad (13)$$

It is evident from (13) that in order to emulate non-zero velocity, the unit cell has to be nonreciprocal, as for a reciprocal symmetric unit cell we have $AD - BC = A^2 - BC = 1$. Furthermore, as the response becomes more nonreciprocal, that is, as $AD - BC$ increases, the normalized velocity $V$ increases logarithmically. Knowing the effective wave impedance and refractive index, the effective permittivity and permeability of the TL can also be easily determined though, in this paper, we limit the analysis mostly to the refractive index and velocity parameter.

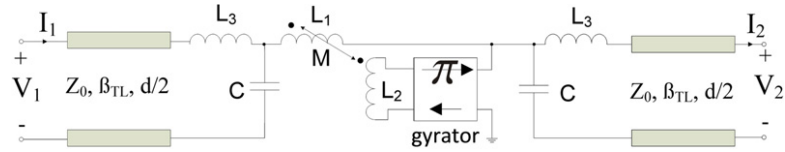

**Figure 2.** Unit cell under study.

## 4. Proposed unit cell

As we need the unit cell to be nonreciprocal, we must utilize some nonreciprocal circuit component. On the other hand, the unit cell also has to be symmetric. Therefore, we suggest the circuit topology shown in figure 2. The nonreciprocity of the circuit is due to a gyrator connected to the TL through coupled inductors. A gyrator is a nonreciprocal two-port circuit component providing 0° phase shift in one direction and 180° phase shift in the opposite direction. The ideal gyrator is passive, linear, and lossless, though the practical realizations are typically active, needing DC-biasing. Even though it is a lossless component, it is characterized by so called gyration resistance $R$, which relates the voltage in port 1 ($V_1$) to the current in port 2 ($I_2$) and vice versa, so that the ABCD matrix of the gyrator has the form

$$\begin{bmatrix} A & B \\ C & D \end{bmatrix}_{\text{gyrator}} = \begin{bmatrix} 0 & R \\ 1/R & 0 \end{bmatrix}. \tag{14}$$

Clearly, this matrix corresponds to a nonreciprocal element, as we have $AD - BC \neq 1$. In the proposed unit cell, the gyrator is used to change the direction of the current flowing from the shunt inductor $L_2$.

Let us first analyze the unit cell of figure 2 without the symmetric series inductors ($L_3$), shunt capacitors ($C$), and TL segments (with length $d/2$, characteristic impedance $Z_0$, and wavenumber $\beta_{TL}$), which are obviously reciprocal and therefore have no effect on the effective normalized velocity $V$. The ABCD parameters of the proposed unit cell can be derived using simple circuit analysis. Considering the equivalent circuit for the coupled inductors (e.g., [11]), we write $V_1$ as a function of $I_1$ and $V_2$ and use the ABCD matrix for the gyrator (14) to write the input/output voltage of the gyrator as a function of its output/input current. Finally, applying Kirchhoff's current law at the node point, we write a set of equations that can be solved for the ABCD parameters (i.e., $V_1$ and $I_1$ as a function of $V_2$ and $I_2$). The ABCD matrix for the unit cell reads

$$\begin{bmatrix} A & B \\ C & D \end{bmatrix}_{\text{uc}} = \begin{bmatrix} 1 - \dfrac{\omega^2 L_1 L_2}{R(R + i\omega M)} - \dfrac{i\omega M}{R} & \dfrac{i\omega L_1 R}{R + i\omega M} \\ \dfrac{i\omega L_2}{R(R + i\omega M)} & \dfrac{R}{R + i\omega M} \end{bmatrix}. \tag{15}$$

As stated earlier, the unit cell should to be symmetric (i.e., $A = D$) in order to ensure a purely moving-media response. By equating the $A$ and $D$ parameters of the proposed unit cell, we get the condition $M = \pm \sqrt{L_1 L_2}$. As the basic definition for mutual inductance is $M = k\sqrt{L_1 L_2}$, where $k$ is the mutual coupling coefficient, we can see that the unit cell is perfectly symmetric when the coupling between the inductors is perfect ($k = \pm 1$ where $\pm$ corresponds to opposite coupling directions between the inductors). However, as long as we can satisfy the condition $R^2 \gg \omega^2(L_1L_2 - M^2)$, the unit cell can be considered approximately symmetric. The other condition for a moving media TL is the nonreciprocity condition $AD - BC \neq 1$. Again, by plugging the equations for the ABCD parameters into this relation, we get the condition $AD - BC = (R - i\omega M)/(R + i\omega M) \neq 1$. This is true always when we have mutual coupling between the two inductors. As the TL segments and additional symmetrically placed loading elements shown in figure 2 do not affect the nonreciprocity or symmetry of the circuit, we can conclude that the suggested unit cell can work as a meta-molecule in an effective moving medium.

Neglecting the series inductors, shunt capacitors, and TL segments, we can write the effective normalized velocity for a cascade of such unit cells using the circuit element values as

$$V = \frac{i}{2k_0d} \ln\left(\frac{R - i\omega M}{R + i\omega M}\right). \tag{16}$$

Furthermore, we can write the effective refractive index as

$$n = \frac{i}{2k_0d} \ln\left(\frac{\frac{R}{\sqrt{(R + i\omega L_1)(R + i\omega\sqrt{L_1L_2})}} - \frac{i\omega\sqrt{L_1L_2}}{R + i\omega\sqrt{L_1L_2}}}{\frac{R}{\sqrt{(R + i\omega L_1)(R + i\omega\sqrt{L_1L_2})}} + \frac{i\omega\sqrt{L_1L_2}}{R + i\omega\sqrt{L_1L_2}}}\right). \tag{17}$$

Notably, the refractive index is independent of the mutual inductance between the coupled inductors, whereas the normalized velocity is not. This allows us to vary the normalized velocity independently of the refractive index by changing the mutual coupling between inductors. On the other hand, by adding symmetric series inductors and/or shunt capacitors (or, alternatively, TL segments) to the unit cell as shown in figure 2, the refractive index can be changed independently of the normalized velocity.

If we further assume that $L_1 = L_2 = L$, (17) simplifies to

$$n = \frac{i}{2k_0d} \ln\left(\frac{R - i\omega L}{R + i\omega L}\right). \tag{18}$$

Clearly, in this case the equation for the refractive index has a similar form to (16), with only the mutual inductance replaced with the coil inductance. Furthermore, if we demand $M = \pm L$, i.e., perfect coupling between the inductors with ± corresponding to opposite coupling directions between the inductors, we can see that in this special case we have, in fact, $n = \pm V$. Looking at (5), this means that for one direction the propagation constant is zero for all frequencies, while for the opposite directions it is non-zero, with a value depending on $R$ and $L$. Unfortunately, such a special case is very difficult to realize for a few different reasons, as becomes evident later.

In order to study the effect of the coupling coefficient $k$ to the effective material parameters, let us consider an example: namely, the unit cell shown in figure 2 with the component values $L_1 = L_2 = 1.15\ \mu\text{H}$, $L_3 = 0$ H, $C = 0$ F, and $R = 365\ \Omega$. Furthermore, let us assume that the period is 0.3 m (electrically small in the studied frequency range) and ignore the effect of the TL segments. First, let us look analytically at the case when we have an ideal gyrator, and the coupling between the inductors changes. The refractive index and the normalized velocity for five different coupling coefficients $k$ are plotted in figure 3. In all the cases, both the refractive index and the normalized velocity are almost constant in the studied frequency range. As was shown before, the refractive index is independent of $k$, while the normalized velocity drops considerably when $k$ is decreased. Even a relatively small decrease of the coupling coefficient from $k = 1$ to $k = 0.9$ (still very strong coupling) causes the normalized velocity to change from 3.14 to 2.83. With a physical coil, we expect the coupling to be about $k = 0.8$. Also, if the coupling coefficient is negative, which corresponds to changing the direction of coupling, the sign of the normalized velocity also changes, as can be seen from (16). This also happens if the whole unit cell is inverted. The Bloch impedance in the ideal case ($k = 1$) is the same for both propagation directions, as expected, and is equal to the gyration resistance. However, this is not the case if we have imperfect coupling between the inductors ($k < 1$). This is illustrated in figure 3(b), which shows the Bloch impedances for coupling coefficients 1, 0.8, and 0.6. As the coupling between the inductors becomes weaker, the magnitude of the imaginary part of the Bloch impedance increases. This imaginary part has equal amplitude but different signs for different propagation directions. This propagation direction dependent part of the Bloch impedance can be attributed to another type of bianisotropic coupling, namely, reciprocal omega coupling [9]. However, in this case, even with the poorest coupling $k = 0.6$, the amplitude of the imaginary part of the Bloch impedances is very small compared to the real part, meaning that the unit cell can be considered practically symmetric; that is, the omega coupling is very weak, especially at low frequencies.

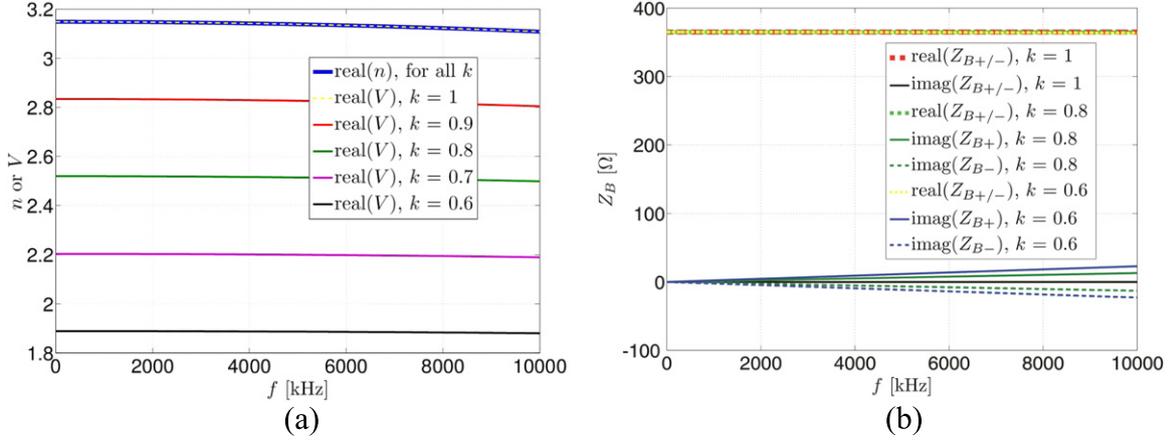

**Figure 3.** (a) Refractive index and normalized velocity for different coupling coefficients in the ideal gyrator case; (b) Bloch impedances for different coupling coefficients in the ideal gyrator case.

In the earlier analysis, the effect of the TL segments, i.e., the host line, was neglected. In practical unit cell implementations, we would always have some finite TL segments connecting the loading elements (the coupled inductors—gyrator circuits) together. For periodical loading with conventional reciprocal passive elements, the effect of the electrically small TL segments is typically negligible, as the loading element itself produces a considerably larger phase shift compared to the TL segments. However, in this case the loading element may provide a very small (or ideally zero) phase shift in one direction, meaning that now the TL segments must be taken into account. In fact, the material parameters of the host line limit the effective refractive index so that we always have $n + V$, $n - V > 1$ for conventional right-handed host lines (in the lossless case, i.e., $n, V \in \mathbb{R}$). This can be understood as follows. In the ideal case with perfect coupling between the inductors when the TL segments are neglected, we have $n = \pm V$; that is, the induced phase shift for one propagation direction is 0. When we introduce the TL segments, we introduce an extra phase shift $\beta_{TL}d$, but the period remains the same. Even if the nonreciprocal circuit provides no phase shift, we have a phase shift due to the TL segments $\varphi_\pm = \beta_{TL}d = k_0 n_{TL} d \geqslant k_0 d$, where $n_{TL}$ is the effective refractive index of the TL. In other words, the TL periodically loaded with the nonreciprocal circuit behaves in this case (for one propagation direction) like an unloaded TL. Therefore, the effective refractive index is equal to the effective refractive index of the TL, which is larger or equal to one. Thus, we have $|\beta_\pm| = |k_0(n \pm V)| \geqslant k_0$, or $n \pm V \geqslant 1$.

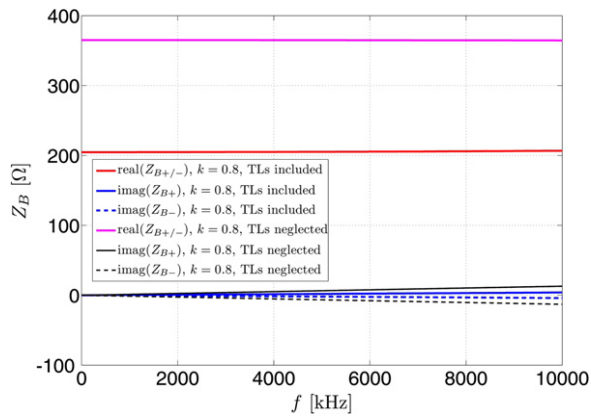

**Figure 4.** Bloch impedances for coupling coefficients $k = 0.8$ in the ideal gyrator case, when the TL segments are neglected and included.

Let us consider the earlier example with the component values $L_3 = 0$ H, $C = 0$ F, $L_1 = L_2 = 1.15\,\mu\text{H}$, $k = 0.8$, and $R = 365\,\Omega$—but now with 0.15 m long TL segments ($\lambda_0/2000$ at 1 MHz) having 50 $\Omega$ characteristic impedance and air-filling on each end of the unit cell. The period of the unit cell is now assumed to be equal to the total length of the two TL segments; that is, the physical length of the loading circuit is considered to be negligible compared to the length of the TLs. The effective refractive index and normalized velocity in this case are calculated using (12) and (13) when the coupling between the inductors is realistic ($k = 0.8$). The refractive index and normalized velocity are practically constant in the studied frequency range (0–10 MHz) and have the values 5.86 and 2.5, respectively. Now, comparing these values to the corresponding cases in figure 3(a), we can see that while the normalized velocity is the same as before, i.e., it still has the form given in (16), the refractive index is clearly increased due to the host line. This makes sense, as the TLs are reciprocal elements, while the parameter $V$ is a measure for the nonreciprocity of the medium. Therefore, the addition of TL segments cannot affect $V$, as any effect the TLs have is reciprocal; that is, it is seen only in $n$. We always have $n + V$, $n - V > 1$ for realistic unit cells with right-handed host TLs, meaning that the (real) propagation constant cannot be smaller than the free space wavenumber. Thus the propagation constant for one propagation direction approaches the free space wavenumber for the air-filled host TL case when the coupling between the inductors approaches the ideal case $k = 1$ and the characteristic impedance is equal to the gyration resistance. The Bloch impedance can also be significantly affected by the inclusion of even electrically short TL segments. The Bloch impedances in the studied case, when the TL segments are neglected and included, are shown in figure 4. Clearly, the amplitudes for the real and imaginary parts of the Bloch impedances for both directions decrease due to the addition of the TL segments. Nevertheless, the Bloch impedance is still practically the same for both propagation directions, meaning that the unit cell can be considered practically symmetric. Decreasing the coupling coefficient was observed, also in this case, to increase the amplitude of the imaginary part of the Bloch impedances (similar to figure 3(b)), thus making the unit cell less symmetric.

Obviously, the amplitude of the coupling coefficient $k$ cannot exceed unity for conventional coupled inductors. Having $|k| > 1$ would imply that power is somehow pumped into the system. However, if we would allow this, an interesting special case would arise. Now, we would no longer be limited by the condition $n + V, n - V > 1$, and the propagation constants could, in fact, have the same sign. This would, in turn, mean that the phase of the wave could only propagate in one direction (as phase velocities $v_p = \omega/\beta_{\pm}$ have the same sign). Moreover, since $n$ and $V$ are practically dispersionless, the group velocities ($v_g = \partial\omega/\partial\beta_{\pm}$) would also have the same sign. In other words, propagation would only be allowed in one direction via two forward wave modes of different propagation constants. The question of how to realize such actively coupled coils is out of the scope of this study.

Previously, we found that the studied unit cell is limited by the condition $n \pm V > 1$, i.e., $|V| < n - 1$, due to its passive nature and the phase shift caused by the the host line. Here, we will discuss how this limitation compares to the case of physically moving magnetodielectric media characterized by (3) and (4). When a magnetodielectric medium (with $n' > 1$) is put into motion, $V$ becomes non-zero. The parameter $|V|$ increases as the velocity of the medium increases (depending on the direction of movement $V > 0$ or $V < 0$), but so does the effective refractive index $n$. As the velocity of the medium approaches the speed of light in the corresponding stationary medium (i.e., $v \to c/n'$), the normalized velocity $|V|$ approaches the effective refractive index $n$ while, simultaneously, both of the parameters approach infinity. Therefore, $|V|$ is always smaller than the effective refractive index of the medium $n$; that is, the parameters $n$ and $V$ are limited by the condition $|V| < n$ when we have $v < c/n'$. However, if we allow the velocity to be between the speed of light in the stationary medium and the speed of light in vacuum (so called Čerenkov zone [2]), the sign of both $n$ and $V$ flips, and we have always $|V| > |n|$. Comparing these two conditions to the condition for the TL unit cell ($|V| < n - 1$), we can see that the effective velocity of the TL realization is always smaller than the speed of light in the corresponding stationary medium ($v < c/n'$). Only if we allow the inductor coupling coefficient to be greater than unity (that is, pumping energy into the system) can we go beyond this limit. However, it should be noted that in the case of a physically moving magnetodielectric medium, parameters $n$ and $V$ cannot be controlled independently as in the TL realization, and when $V$ has a value close to $n$, both of these values are always very high.

One key difference between a physically moving medium and an artificial moving medium has not yet been addressed. As was mentioned in the Introduction, a wave propagating from a stationary medium to a moving medium experiences a Doppler shift due to the movement of the interface between the two media. The same is true for a wave reflected from such an interface. However, in the artificial moving medium unit cell, no such effect can occur due to the linear nature of the circuit elements comprising it (i.e., the generation of new frequency components, not present in the spectrum of the excitation signal, is not possible in linear systems). Therefore, our model when considering a finite slab corresponds, in fact, to a medium which is moving, but its boundaries remain stationary, similar to the movement of a conveyor belt (the other side of which is hidden from view). Obviously, this is also the case for any artificial realization of moving media lacking nonlinear elements or materials (e.g., [3, 5, 8]).

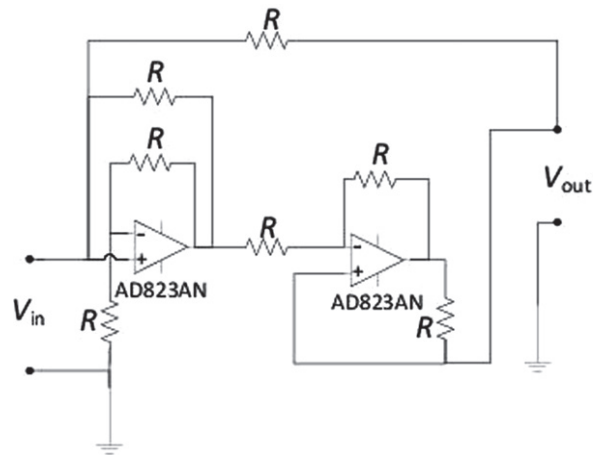

**Figure 5.** Chosen gyrator realization (op amp biasing and DC decoupling capacitors are left out for clarity).

## 5. Experimental results

Our goal was to manufacture and experimentally study the unit cell analyzed analytically and numerically above (figure 2) with the component values $L_3 = 0$ H, $C = 0$ F, $L_1 = L_2 = 1.15$ μH, and $R = 365$ Ω, with 0.15 m long TL segments with the 50 Ω characteristic impedance and air-filling on each end of the unit cell ($d = 0.3$ m). For this proof-of-concept study, the operational frequency was limited to approximately 1 MHz, as the chosen gyrator circuit was observed in simulations to provide the needed phase shift only below approximately 1 MHz.

In order to realize the unit cell of figure 2 with the aforementioned component values, we first had to choose the gyrator. A myriad of different circuits can be used for realizing a gyrator [15]. We have chosen to use the circuit shown in figure 5, which is based on two operational amplifiers (AD823). The coupled inductors in the unit cell were realized as two coils wound on top of each other (separated by a piece of tape), with nine turns each around a 12 mm plastic core. The diameter of the used copper wire was 0.5 mm. The values of the inductors and the strength of the coupling between them were measured to be approximately $L_1 = 0.97$ μH, $L_2 = 0.805$ μH, and $k = 0.835$ in the frequency range 300 kHz–1 MHz. The ohmic losses in the inductors were measured to be approximately 0.18 Ω and 0.06 Ω for inductors $L_1$ and $L_2$, respectively. As the goal is simply to demonstrate the concept, the slight discrepancy between the actual measured values and the intended values is of little importance. For practical reasons, the 0.15 m long TL segments (air-filled) were replaced with 0.1 m long pieces of 50 Ω coaxial cable RG-58/U ($\epsilon_r = 1.3$). The DC bias voltage was $\pm 6$ V. The manufactured unit cell is shown in figure 6.

The S-parameters (e.g., [10]) for the unit cell were measured using a Rohde & Schwarz ZVA8 vector network analyzer (VNA). By converting these into ABCD parameters, the dispersion, Bloch impedances, and material parameters for an infinite cascade of such unit cells can be extracted using the equations derived earlier. These results were compared to the numerical results based on S-parameters simulated using Agilent Technologies' Advanced Design System (ADS). In the numerical model, the operational amplifier is modeled using a Simulation Program with Integrated Circuit Emphasis (SPICE) model provided by the manufacturer, and the inductors have values corresponding to the manufactured coils with realistic losses ($L_1 = 0.97$ μH, $L_2 = 0.805$ μH, $k = 0.835$, $R_{L1} = 0.18$ Ω, and $R_{L2} = 0.06$ Ω).

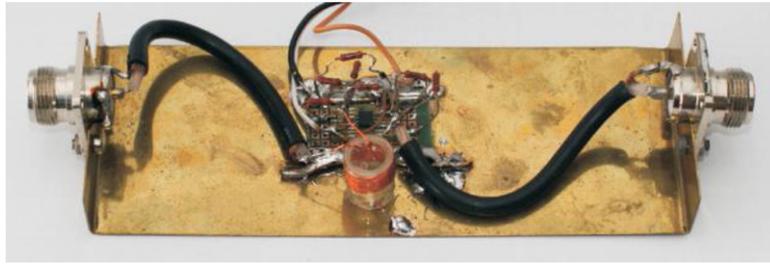

**Figure 6.** Manufactured unit cell.

The effective refractive index $n$, the effective normalized velocity $V$, and the dispersion diagram of the media (having the unit cell shown in figure 6 calculated based on measurements and ADS simulations) are presented in figures 7(a)–(c). Clearly, the unit cell operates as a moving media TL, as $V$ has a non-zero value, and the amplitudes of the propagation constants are different for different propagation directions. However, the non-ideal gyrator with imperfect phase-shifting, losses, imperfect matching, and non-zero $A$ and $D$ parameters for the gyrator (which allow part of the signal to pass through the gyrator without the intended nonreciprocal phase shift) considerably limits the performance of the circuit, roughly doubling the refractive index compared to the ideal case with a perfect gyrator and lossless elements, which was analyzed in section 4. The normalized velocity remains almost the same. As the propagation constant is proportional to the sum or subtraction between $n$ and $V$, depending on the propagation direction, the unit cell is, therefore, effectively less nonreciprocal than expected. Also, some small differences can be seen between the simulated and measured results. The primary reason for this discrepancy lies in the high measurement uncertainty of the VNA at very low frequencies. Furthermore, the gyrator circuit as well as the coils have some small intrinsic losses, which can be seen as the imaginary parts of the refractive index and the normalized velocity.

The Bloch impedances for different propagation directions are shown in figure 7(d). Again, the measured results differ slightly from the simulation results for the aforementioned reasons. Though the unit cell with a perfect gyrator and perfect coupling between inductors is symmetric, the manufactured unit cell is slightly asymmetric due to the non-ideal symmetry in the component losses and imperfect coupling. This can be seen in the Bloch impedances for different propagation directions. While the real parts of the Bloch impedances are practically the same for different propagation directions, the difference is more considerable in the imaginary parts. However, as the real parts are much larger than the imaginary parts, the omega coupling effect is still quite small. This small asymmetry was taken into account when plotting the dispersion curves by using the more general form for the dispersion equation (10). Moreover, the Bloch impedances are notably much smaller than the ones calculated for the ideal unit cell (figure 4). This is, again, mostly due to the discrepancy between the ideal gyrator model and the circuit implementation.

Earlier, we extracted the material parameters of the unit cell ($n$ and $V$) as defined by the constitutive relations given in (1) and (2), that is, the material parameters for the special class of nonreciprocal media called moving media. On the other hand, these parameters are connected to the material parameters of real magnetodielectric media in motion through (3) and (4). Here, we will study which kind of physically moving magnetodielectric medium would correspond to our artificial unit cell. As we have, according to figures 7(a) and 7(b), $|V| < n$ at all frequencies, we can expect the velocity of the corresponding physically moving magnetodielectric medium to be smaller than the speed of light in the corresponding stationary medium. Knowing the effective refractive index *n* and velocity parameter *V*, (3) and (4) can be solved numerically for the refractive index of the medium $n'$ as well as the speed of the medium *v*. Naturally, permittivities and permeabilities ($\epsilon$, $\mu$, $\epsilon'$, and $\mu'$) also could be easily derived, but here we, again, limit our analysis to the refractive indices. The resulting refractive index and velocity for the measured unit cell of figure 6 are shown in figure 8. Noticeably, the refractive index is almost equal to the effective refractive index plotted in figure 7(a) (though slightly lower). This tells us that the effect of special relativity is fairly small; i.e., the term $(1 - \frac{v^2}{c^2})/(1 - \frac{n'^2 v^2}{c^2})$ is fairly close to unity. This assumption is confirmed by looking at the speed of the medium, which is always about two orders of magnitude smaller than the speed of light in vacuum and one order of magnitude smaller than the speed of light in the stationary magnetodielectric media. Notably, the velocity of the medium *v* also has a considerable imaginary part which, though non-physical in the context of realizing real moving magnetodielectric media, opens the door for realizing exotic media not found in nature. As described in [6], this imaginary part induces losses for waves propagating in one propagation direction while providing gain for waves propagating in the opposite direction. The energy for this amplification comes, in this case, from the operational amplifiers of the gyrator or, to be precise, from the DC power supply providing a biasing voltage for them. However, the reciprocal losses in the medium are larger than this nonreciprocal loss/gain, meaning that the medium altogether acts as a lossy medium for both propagation directions, albeit with the amount of losses being different for different propagation directions, as can also be seen in figure 7(c).

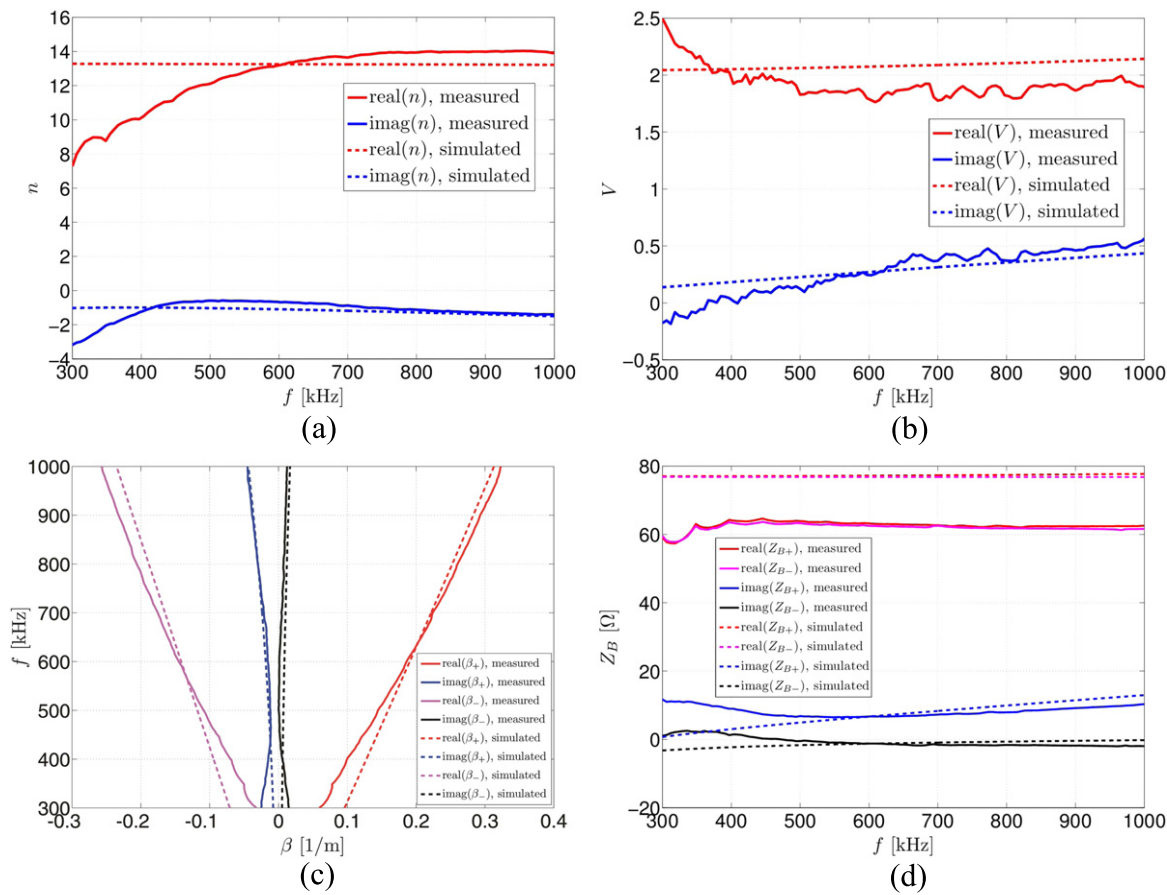

**Figure 7.** Measurement and simulation results for the unit cell shown in figure 6: (a) effective refractive index *n*, (b) normalized velocity parameter *V*, (c) dispersion in an infinite cascade of unit cells, and (d) Bloch impedance.

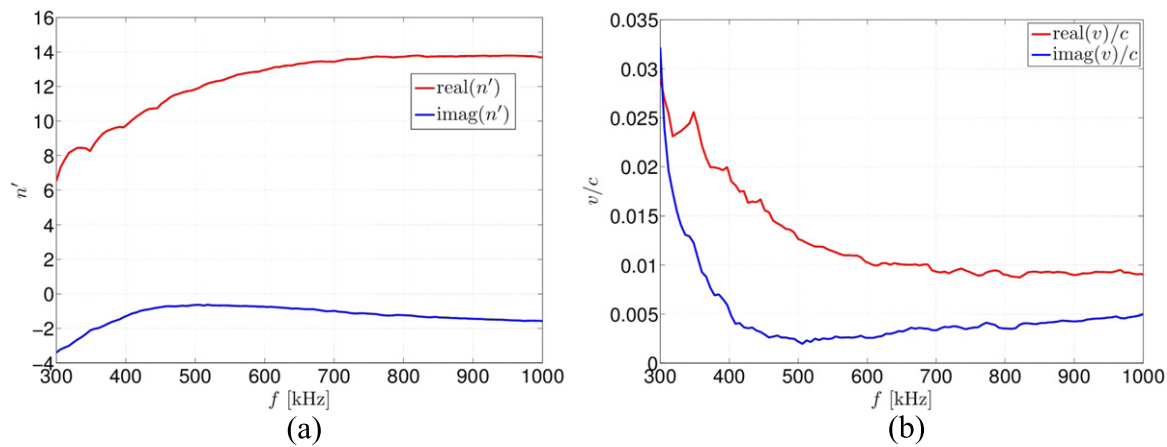

**Figure 8.** Refractive index $n'$ (a) and velocity $v$ (b) of the moving magnetodielectric medium, which is emulated by the artificial unit cell shown in figure 6.

## 6. Moving medium with composite right-/left- handed host line

Earlier, it was assumed that the permittivity and permeability of the medium were both positive; i.e., the host line was a conventional right-handed transmission line. In that case, the refractive index of the host line limited the dispersion so that we always had $n + V$, $n - V > 1$ leading to the limits $\beta_+ > k_0$ and $\beta_- < -k_0$, according to (5). It is well known that an ideal left-handed transmission line (characterized by an equivalent circuit consisting of a series capacitor and a shunt inductor) has a negative refractive index [13]. Therefore, in order to go beyond the aforementioned limits, we will, here, consider the case where the host line is left-handed (LH) or composite right-/left- handed (CRLH), to be precise. A physical TL with series capacitor and shunt inductor loading always corresponds to a composite right-/left- handed transmission line (CRLH TL), meaning that in addition to the series capacitance and shunt inductance, there also exists parasitic effective series inductance and shunt capacitance due to the host line. Such TLs can support both forward and backward waves as opposed to pure right- or left-handed TLs, which can support only forward or backward waves, respectively. Backward waves have the unique property of having opposite phase and energy velocities. However, the response of a CRLH TL is always reciprocal, meaning that the same propagation mode is supported for both propagation directions at a given frequency. In [12], the concept of nonreciprocal CRLH TLs was discussed. It was shown that such TLs have the ability to support forward waves for one (power) propagation direction and backward waves for the opposite direction at the same frequency. However, in that paper, nonreciprocity was achieved by employing a ferrite substrate with a biasing magnetic field. In other words, the nonreciprocity was due to the nonreciprocal permeability dyadic of the medium as opposed to any bi-anisotropic response. In contrast to the moving medium, in this case the (Bloch) impedance is different for different propagation directions, as a ferrite-loaded TL is an asymmetric element.

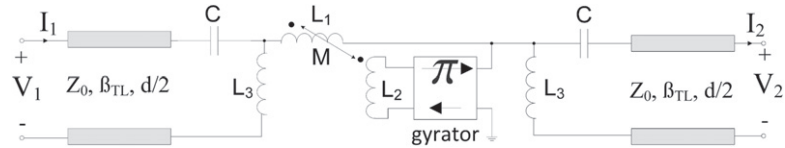

**Figure 9.** Unit cell with CRLH host line.

Let us consider the unit cell analyzed earlier with one modification: instead of having a series inductor and a shunt capacitor placed symmetrically on both sides of the core inductor-gyrator circuit as seen in figure 2, we have a series capacitor and a shunt inductor as seen in figure 9. The gyrator is considered to be ideal with gyration resistance of 100 Ω. As before, we have $L_1 = L_2 = 1.15\,\mu$H and $k = 0.8$, and the air-filled TL segments are 0.15 m long with 100 Ω characteristic impedance. The additional components have the values $C = 100$ nF and $L_3 = 800$ μH. Again, the total response of the unit cell (i.e., total ABCD matrix) can be easily calculated analytically using known ABCD matrices for individual circuit elements. Knowing the ABCD parameters for the whole cascade, we can calculate the dispersion, Bloch impedance, refractive index, normalized velocity $V$, permittivity, and permeability for the given unit cell using equations derived earlier.

The dispersion diagram of the proposed unit cell is shown in figure 10(a). At low frequencies, only a backward wave can propagate in the structure with the propagation constant having different values for opposite propagation directions. However, at frequency 509 kHz the propagation constant corresponding to the positive (energy) propagation direction changes sign, meaning that the phase velocity also changes sign. Therefore, at frequencies 509 kHz–638 kHz as well as 711 kHz–1 MHz, only a forward wave solution is possible for a wave propagating in the positive direction, while for the opposite direction only a backward wave can propagate at the same frequencies. At frequencies 638 kHz–711 kHz, there exists a stopband (independent of the propagation direction); i.e., the CRLH TL is not balanced in this case. By changing the values of the series capacitors and/or shunt inductors (e.g., changing $C$ from 100 nF to 80 nF), the balanced condition could be achieved and thus the stopband eliminated. While the phase velocity of the other solution changes sign at 509 kHz, the group velocity has a different sign for the two solutions at all frequencies, unlike in the aforementioned case where the host line was a common right-handed TL, but the coupling coefficient between the inductors was allowed to exceed unity (implying activity). This is due to the dispersion of $n$, which is an unavoidable property of all passive media with a negative refractive index. Having a dispersionless and negative refractive index for a TL unit cell is only possible if active non-Foster elements (negative capacitors/inductors) are used [14].

It should be noted that, here, backward propagation does not imply that the effective refractive index, permittivity, and permeability are negative (as in a typical reciprocal CRLH media) due to the effect of the velocity parameter $V$. This is demonstrated in figures 10(b) and 10(c), where the effective refractive index and velocity parameter as well as effective permittivity and permeability are plotted, respectively. While at the lowest studied frequencies the refractive index as well as both permittivity and permeability are negative, at frequencies higher than 700 kHz, all of them have positive values. However, despite this, the forward wave exists for only one of the propagation directions due to non-zero velocity parameter *V*, as can be seen from the dispersion diagram of figure 10(a). It should be noted that, as before, *V* depends on the gyration resistance, the strength of the coupling between the inductors, and the period of the structure, but not on the properties of the host TL. By changing the strength of the coupling (i.e., coupling coefficient *k*), *V* can be changed independently of the other material parameters. Notably, when only one of the material parameters (permittivity or permeability) is negative, we naturally get an imaginary refractive index which leads to a non-zero imaginary part of the propagation constant (i.e., a stopband). Nevertheless, the real part of the propagation constant is non-zero as is also its slope, again due to the velocity parameter *V*. In the case of a physically moving magnetodielectric medium with $n' > 1$ or $n' < -1$, we are, according to (3) and (4), limited by the condition $|V| < |n|$ when the velocity of the medium is smaller than the speed of light in the stationary medium and $|V| > |n|$ when it is larger than that. Clearly, the first condition is satisfied only at the lowest analyzed frequencies, meaning that extremely high velocities of a real magnetodielectric medium are needed in order to replicate the response at higher frequencies.

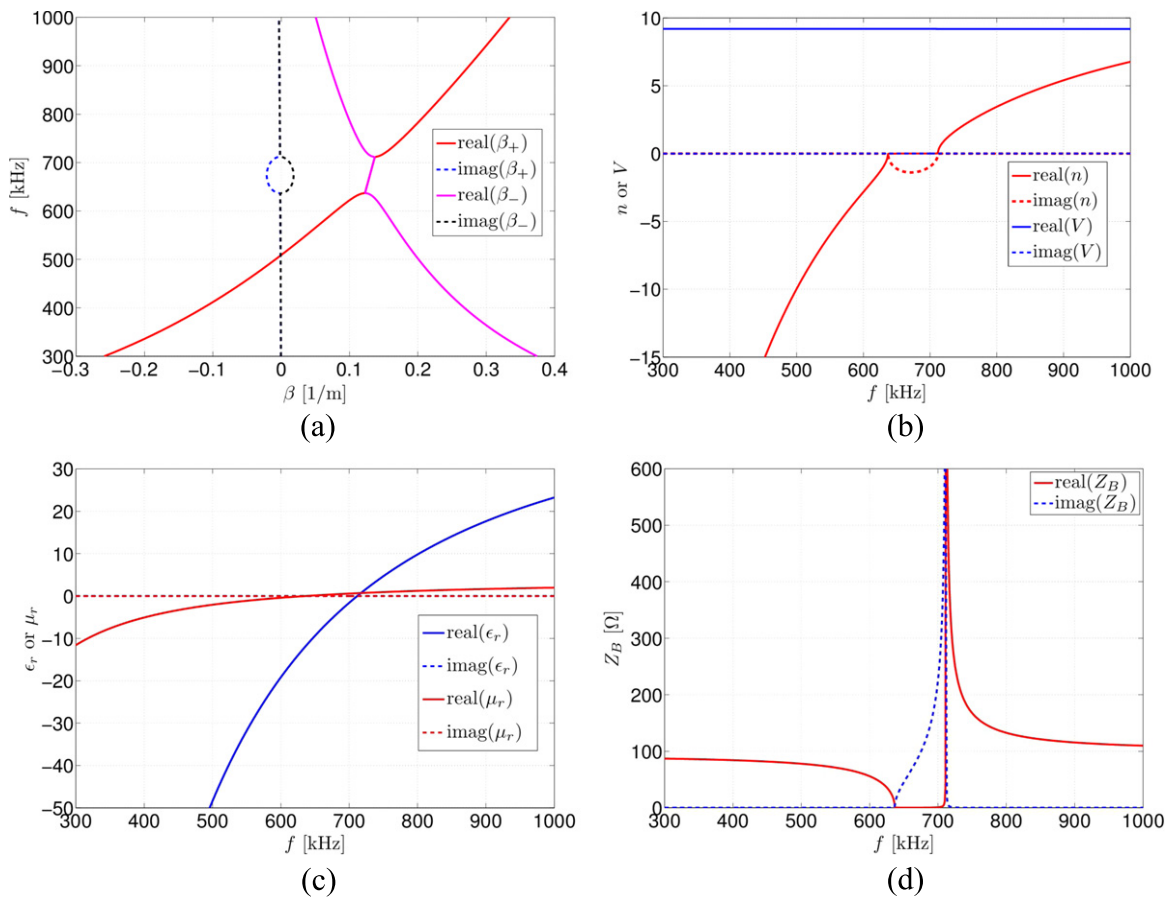

**Figure 10.** Simulation results for for the unit cell shown in figure 9: (a) dispersion in an infinite cascade of unit cells, (b) effective refractive index *n* and normalized velocity parameter *V*, (c) effective permittivity and permeability, and (d) Bloch impedance.

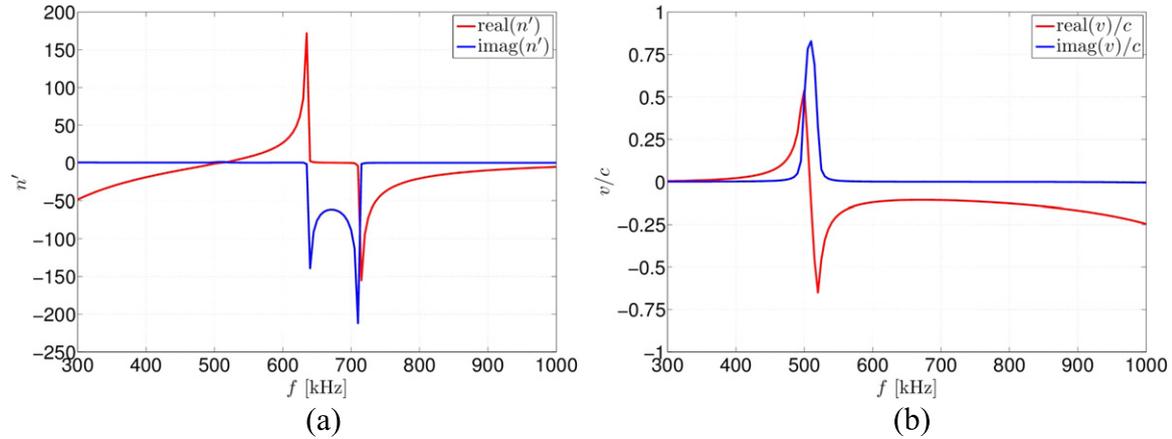

**Figure 11.** Refractive index $n'$ (a) and velocity $v$ (b) of the moving magnetodielectric medium, which is emulated by the artificial unit cell shown in figure 9.

The Bloch impedance in this case is the same for both propagation directions and is shown in figure 10(d). The real part of the Bloch impedance is zero in the stopband and fairly constant in the two passbands (though having a resonance peak between the stopband and the second passband), while the imaginary part is zero in the passband and non-zero and increasing in the stopband. If the CRLH TL were balanced, the Bloch impedance would be almost constant, as the stopband would disappear.

Finally, the refractive index of the real magnetodielectric medium in motion corresponding to the artificial unit cell shown in figure 9 and the corresponding velocity of the medium are shown in figures 11(a) and 11(b), respectively. These were acquired, as before, by solving (3) and (4) for $n'$ and $v/c$ numerically. The behavior for the unit cell with the CRLH host line is quite exotic. At the lowest examined frequencies, the refractive index is negative and increasing while the velocity is positive and increasing. As the refractive index approaches zero, the velocity approaches the speed of light in vacuum. Finally, at 510 MHz the velocity has a resonance! This is the point where we have $n = -V$ and thus $\beta_+ = 0$ 1/m. For the rest of the analyzed frequencies, the velocity is negative; i.e., the direction of movement changes. These higher frequencies (excluding the stopband) correspond to the Čerenkov zone where the velocity of the medium exceeds the speed of light in the corresponding stationary medium ($c/n'$). However, the real part of the velocity still never exceeds the speed of light in vacuum. The refractive index has a negative sign not only at 300 MHz–510 MHz, where only backward waves are supported, but also at 715 MHz–1000 MHz, where both backward and forward waves are supported. On the other hand, the sign of $n'$ is positive at 510 MHz–635 MHz, where both backward and forward waves also are supported. In the stopband (510–715 MHz), $n'$ is purely imaginary. Also, the slope of the real part of the refractive index is always positive as expected for passive media.

In the previous example, the physical velocity of the medium $v$ was always smaller than the speed of light in vacuum. However, this is not always the case. Looking at (3) and (4) and demanding $v/c > 0$, we can see that this means, in view of the effective parameters $n$ and $V$, that we must simply have $|n|, |V| < 1$ (assuming $n' > 1$ or $n' < -1$). This can be achieved, for example, by taking the unit cell analyzed previously in this section and reducing the value of the inductors $L_1$ and $L_2$ from 1.15 μH to 4.15 nH. The material parameters for this case are shown in figure 12. Clearly, the velocity exceeds the speed of light in vacuum on both sides of the stopband, appearing approximately at 1.85 MHz–2.17 MHz. However, the effective refractive index is notably highly dispersive here, which means that, as for the previous example, the phase of the wave propagates only in one direction at some frequencies, but the energy can still propagate in both directions at all frequencies.

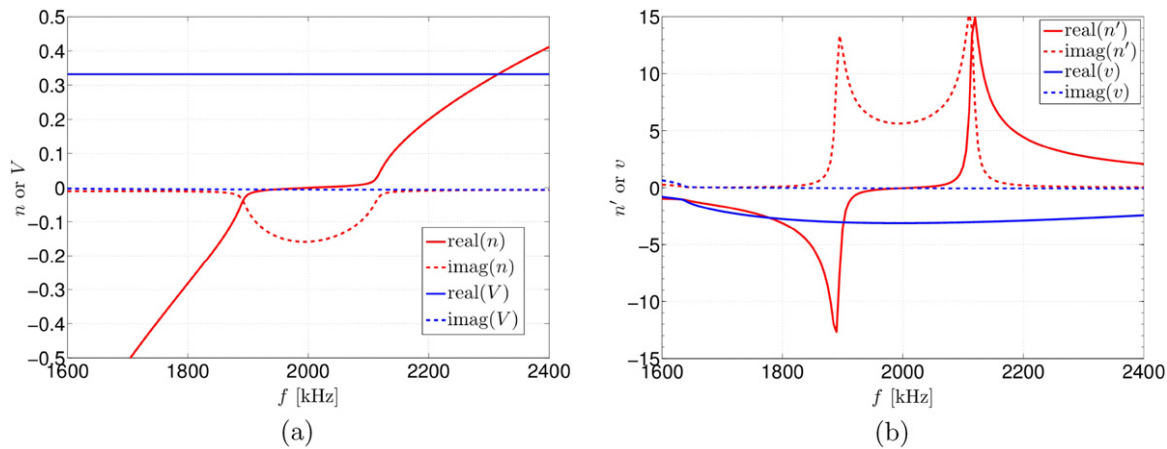

**Figure 12.** Effective material parameters *n* and *V* (a) and the physical material parameters *n'* and *v* (b) of the moving magnetodielectric medium, which is emulated by the artificial unit cell shown in figure 9 in the case where the circuit parameters are chosen so that physical velocity can exceed the speed of light in vacuum.

## 7. Discussion

We have presented a realization of an artificial moving medium, that is, a nonreciprocal medium at rest that effectively mimics the response of a magnetodielectric medium in motion. It should be noted that though nonreciprocal media (both volumetric and transmission-line-based) have been demonstrated before, this is to the best of our knowledge the first time that a medium with a purely moving medium type of coupling has been demonstrated. The design is based on transmission lines periodically loaded with symmetric, nonreciprocal circuits utilizing a gyrator. The proposed unit cell has been analyzed analytically, numerically, and experimentally and has been shown to behave effectively as a moving medium in a wide frequency range. Though there is some omega coupling present in the proposed unit cell, as evidenced by the Bloch impedance changing slightly with the propagation direction, it is very weak. In other suggested realizations ([3, 5, 8]), the effect of other bianisotropic coupling phenomena (chiral coupling, omega coupling) is much more prominent. It was shown that the velocity parameter *V* can be varied without affecting the effective refractive index of the medium by simply changing the coupling between two inductors. The experimental results suggest that realizing complex velocity is also possible using the proposed unit cell, opening a door for realizing novel nonreciprocal devices not possible even by utilizing actually moving media. In the studied case, the effective velocity of the medium was considerably smaller than the speed of light in the corresponding stationary medium. This is always the case for the proposed unit cell, as the loading circuit is passive (though realizations of gyrators are typically active). However, with a truly active loading circuit, the speed of the medium is not limited, and velocities larger than the speed of light in the corresponding stationary medium (and possibly even the speed of light in vacuum) could be achieved. In such media, energy could propagate only in one direction. As always with active circuits, stability conditions for the circuit must, in that case, be taken into account, which complicates the design. While the prototype shown here was designed for low radio frequencies, the concept can be extended to any frequency range, though practical realization of the unit cell is more difficult at higher frequencies. The state-of-the-art high-speed operational amplifiers can have bandwidths up to a few gigahertz, though a novel solution would be needed in order to realize mutually coupled inductors accurately at such high frequencies. Finally, it was shown that using a CRLH host line instead of conventional right-handed host line in the proposed unit cell allows us to tune the value of the effective refractive index more freely without affecting the velocity parameter $V$. This also allows the realized TL medium to have the unique property of supporting both forward and backward wave propagation at the same frequency, depending on the propagation direction. Such a property could be advantageous in, for example, design of leaky-wave antennas. Furthermore, it was shown that physically moving a magnetodielectric medium corresponding to this artificial unit cell has very exotic refractive index and velocity values not easily realizable with any physically moving magnetodielectric media. Medium velocities exceeding even the speed of light in vacuum can in this case be achieved. However, the material parameters are always highly dispersive for such unit cells due to the negative effective refractive index, meaning that the energy can still always propagate in both positive and negative propagation directions. Again, only by making the unit cell active can we overcome this limitation.

## Acknowledgements

The authors thank Mr. D Petricevic for help in manufacturing the prototype.